\documentstyle[epsf,12pt]{article}

\begin{document}
\thispagestyle{empty}
\newcommand{\vsl}{v\!\!\!/}
\renewcommand{\thefootnote}{\alph{footnote}}
\begin{flushright}
{\sc BU-HEP} 95-5\\
hep-ph/9502370\\
January 1995
\end{flushright}
\vspace{.1cm}
\begin{center}
{\LARGE\sc Numerical investigations of baryon non-conserving
processes in the electroweak theory $^{*}$}\\[2cm]
{\sc Claudio Rebbi and Robert Singleton, Jr.}\\[1.5cm]
{\sc Physics Department \\[3mm]
Boston University\\[3mm]
590 Commonwealth Avenue\\[3mm]
Boston, MA 02215, USA}\\[1.5cm]
{\em * Based on a lecture given by C. Rebbi at the 1994 ICTP
Summer School  in High Energy Physics and Cosmology, to
be published in the proceedings of the School (World Scientific
Publishing - Singapore -
1995).}\\[1.5cm]
{\sc Abstract}
\end{center}
\begin{quote}
We illustrate the use of computational methods
for the study of baryon non-conservation in high energy
electroweak processes.
\end{quote}

\section{Introduction}
\hspace{1cm}
 Since the advent of the modern scientific method,
knowledge of the
physical world has advanced through the interplay of experimental
observation and theoretical analysis.  Correspondingly, the terms
experimental physics and theoretical physics have been used to
characterize these two basic methodologies of scientific research.
The introduction of computers, however, has lead to the
establishment of a third fundamental methodology and the term
computational physics has been coined to denote all those
investigations where computers are used in an intrinsic manner
to unravel properties of complex systems.
The methods of computational physics do not offer a replacement
for either experimental investigation or theoretical analysis.
Indeed, although computer modeling shares many of  features
of actual experiment, clearly the study of computer simulated
phenomena can never replace the observation of the actual
world.  In a more subtle way, computational physics cannot
displace theoretical physics either.  The
insights into the phenomena provided by the theoretical analysis
constitute the necessary platform over which a computational
investigation can be launched and are ultimately required to
interpret the latter's results.  Thus theoretical physics and
computational physics complement each other and, at their best,
proceed hand in hand.

Because of the large computer power required by some numerical
investigations, it is frequently thought that computational physics is
accessible only to a limited constituency of researchers fortunate
enough to have access to powerful supercomputers.  This belief is
wrong, on two accounts.  First, not all computational
investigations require very large resources.  In many instances,
ingenuity more than powerful hardware is the key to a successful
utilization of the computer.  Second, the amazingly rapid pace of
technological development in the field of computers is making these
tools ever more accessible on a world wide scale.  For these reasons,
the
International Center for Theoretical Physics has
been offering colleges on computational physics on a rather
regular basis since the first conference and college it held in
1986.

In particle physics some of the most successful computational
applications have taken place in the area of lattice gauge theories,
where computer simulations of quantum fluctuations have been
used to evaluate several non-perturbative quantities of primary
importance. In particular, computational investigations of QCD
have provided evidence of quark confinement as well as the
means of calculating hadronic observables such as meson
and baryon masses, some weak matrix elements, the deconfining
temperature and the value of $\alpha_S$. However, the span of
computational particle physics investigations is by no means
limited to lattice QCD, and in these lectures we will illustrate
another challenging application of numerical
techniques to particle phenomena where computational and
theoretical methods complement each other.  We will see, indeed,
that the theoretical analysis of the phenomenon provides the basis
for the computational investigation, which in turn can produce
information that goes well beyond the reach of purely analytic
techniques.

The problem we will study is the possible occurrence of baryon
number violation in high energy electroweak processes.  In
perturbation
theory, baryon number is strictly conserved. But, as has been
well  known
since the pioneering work of 't Hooft\cite{thooft76},
the axial vector
anomaly implies that baryon number is not  conserved in processes
which change the topology of the gauge fields. Baryon number
violating amplitudes are non-perturbative, and viable methods
of calculation are scarce. Basically there are two ways of getting
at non-perturbative information in quantum field theory. One can
use either semi-classical techniques or direct lattice simulations
of the quantum fluctuations. Unfortunately, theories with small
coupling constants are not suited for the latter, so the electroweak
sector of the standard model lies beyond the reach of direct lattice
calculations. This means that semiclassical  methods presently
offer the only way to study baryon number violating electroweak
processes.

Electroweak baryon number violation is associated with topology
change of the gauge fields. Classically, gauge field configurations
with different topology (i.e differing by a topologically non-trivial
gauge transformation) are separated by an energy barrier.  The
(unstable) solution of the classical equations of motion which lies
at the top of the energy barrier is called  the sphaleron\cite{KM}.
At energies lower than the sphaleron energy, topology changing
transitions, and hence baryon number violation, can only occur
via quantum mechanical tunneling.  Under certain circumstances,
semiclassical methods can be used to approximate these tunneling
rates. The relevant solution of  the  Euclidean equations of motion
which describe such tunneling is known as the instanton\cite{bpst}.

All transitions that change topology involve fields of order
$1/g$ and actions $ S = S_0/g^2 $, where $S_0$ is a coupling
constant independent action reexpressed in terms of rescaled fields
$ g A_{\mu}(x) $.  As a consequence, all tunneling amplitudes contain
a barrier penetration factor $\exp(-S_0/g^2)$:
\begin{equation}
A \sim \exp(-S_0/g^2),
\label{eq1}
\end{equation}
where $S_0$ is a numerical factor of order one.
The appearance of the square of the electroweak coupling
constant $g$
in the denominator of the exponent in Eq.~\ref{eq1} has three
important implications.  First  it tells us that the phenomenon
is non-perturbative, in the sense that $A$ has an essential singularity
at $g=0$ and thus cannot be expressed in terms of a perturbative
expansion in powers of $g$.  Second, the fact that the actual numerical
value of $g$ is very small indicates that non-perturbative semiclassical
methods, based on saddle point expansions of path integrals around
solutions of the classical equations of motion, are likely to
produce reliable results (to better understand this point, remember
that in units in which $\hbar$ is not set equal to $1$,  Eq.~\ref{eq1}
takes the form $A \sim \exp[-S_0/(\hbar g^2)]$ so that small $g$ and
small $\hbar$ are equivalent).  Third, the small value of $g$ also
makes these processes apparently irrelevant because the associated transition
rates, proportional to  $|A|^2$, turn out
to be abysmally small.

This state of affairs changed a few years ago when Ringwald
\cite{ring} and later Espinosa\cite{esp}
noticed that a summation
of the semiclassical amplitudes over final states gives rise
to factors which increase very rapidly with increasing energy.
This might lead to a compensation of the suppression factor
in Eq.~\ref{eq1} for energies approaching the energy of the
barrier, i.e. the sphaleron energy $E_{sph}$. Intuitively, one
might expect the tunneling suppression factor to become much
less severe as the energy approaches  the energy of the barrier.
In particular one might expect it to disappear altogether for
$E>E_{sph}$, i.e. in the region where the topology changing
processes are classically allowed. Investigations
have indeed confirmed that this is precisely what happens in
high temperature electroweak processes\cite{highT} . As the
temperature approaches $E_{sph}$ the barrier-penetration
suppression factor becomes progressively less pronounced,
and electroweak baryon number violation becomes
unsuppressed  altogether for temperatures comparable
to the sphaleron energy.  The situation is, however, much
less clear for high energy collisions. Phase space
considerations are more
subtle, and simply because one has enough energy to pass
over the barrier does not guarantee that one does.  The problem
is that in high energy collisions the initial state is an exclusive
two particle state, which is difficult to incorporate in a
semiclassical treatment of the transition amplitude.

A possible remedy to this situation has recently been proposed
by Rubakov, Son and Tinyakov\cite{rst}, who suggested
that one considers inclusive initial coherent states, but
constrained so that energy and particle number take fixed
average values
\begin{equation}
E = {\epsilon \over g^2},
\label{eq2}
\end{equation}
\begin{equation}
N = {\nu \over g^2}.
\label{eq3}
\end{equation}
In the limit $g \to 0$, with $\epsilon$ and $\nu$ held fixed, the
path integrals giving the transition amplitudes are then dominated
by a saddle point configuration which solves the classical
equations of motion.  This permits a semiclassical calculation
of the transition rates.  Information on the high energy
collision processes can then be obtained from the limit $\nu \to
0$. While this limit does not strictly reproduce the exclusive
two-particle
initial state, under some reasonable assumptions of continuity
it can be argued that the corresponding transition rates will be
equally suppressed or unsuppressed.

In these lectures we will not reproduce the derivation of the
saddle point equations, which would  form the topic of
an extended set of lectures in itself.  (Indeed, Rubakov
presented such lectures at the 1992 ICTP Summer School
in High Energy Physics and Cosmology, and the conversations
that one of us held with him then stimulated the investigation
we describe here.)
Rather, we will start from these equations, which we will of
course recapitulate, and describe the computational techniques
used to solve them and the progress we
have made in this direction.  For the actual derivation of the
equations the reader should consult Ref. \cite{rst}.

In the next section we illustrate the general properties of topology
changing evolution of the classical fields.  For simplicity we
first consider  the 2-dimensional Abelian Higgs model. Then we
examine  the 4-dimensional $SU(2)$ Higgs model, but restricted
to the spherical {\it Ansatz} to obtain a computationally
tractable system.  In Section 3 we investigate the properties of
topology changing processes above the sphaleron barrier,
i.e. in the classically allowed energy domain (see also
Ref. \cite{rs}). And in Section
4 we  finally describe the equations introduced by Rubakov,
Son and Tinyakov\cite{rst} and the computational methods
required to solve them.

\section{Topology changing field evolution}
\hspace{1cm}
The 1+1 dimensional Abelian Higgs system is defined
in terms of a complex matter field $\phi(x)$ and an
Abelian gauge potential $A_{\mu}(x)$ with  action
\begin{equation}
S = \int dx ^2 ~ \left\{- {1 \over 4} F_{\mu \nu} F^{\mu \nu} +
(D_{\mu}  \phi)^* D^{\mu} \phi  - \lambda (|\phi|^2 -1 )^2
\right\} \ ,
\label{eq4}
\end{equation}
where the indices run over 0 and 1,
$F_{\mu \nu} = \partial_{\mu} A_{\nu} - \partial_{\nu} A_{\mu}$,
$D_{\mu}\phi = \partial_{\mu}\phi - \imath A_{\mu} \phi$ and
many inessential constants have been eliminated by a suitable
choice of units.

The most important feature of this system is that the vacuum,
i.e. the configuration of minimum energy, occurs for non-vanishing $\phi$,
indeed, with our special choice of units for $|\phi| =1$.
Since this does not specify the phase of $\phi$, there is not
a unique vacuum state, but rather multiple vacua.  Still,
because of gauge invariance one must be careful in regard
to the physical significance of the phase of $\phi$. A local
variation of the phase of $\phi$ can always be undone by
a suitable gauge transformation. And since gauge
equivalent configurations must be considered  physically
indistinguishable, local variations of the phase of the matter
field do not lead to different vacua.  However, variations
of the phase of $\phi$ by multiples of $2 \pi$ (as the coordinate
$x^1$ spans the entire spatial axis) cannot be undone by a
local gauge transformation, and thus define topologically
distinct vacuum states.  These vacua differ by the global
topological properties of the field configuration.  The condition
$|\phi|=1$ restricts the values of the matter field to the unit
circle (in the complex plane).  If we demand that $\phi$ takes
fixed identical values as  $x^1 \to \pm \infty$  (a condition we
later relax), then the number of times $\phi$ winds
around the unit circle as $x^1$ spans the entire real axis is a
topological invariant (the winding number) and characterizes
different topologically inequivalent vacuum states.

Figures 1a-c illustrate three contours traced in  the complex
plane by the field variable $\phi(x^1)$ as the coordinate $x^1$
spans the entire space axis. Inequivalent vacuum configurations
with winding numbers 0  and 1 respectively are depicted in
Figs.~1a and 1c.  In the contour of Fig.~1a the phase of
$\phi$ stays fixed at zero as $x^1$ ranges  between  $-\infty$
and $+\infty$,  whereas it goes once around the unit circle
in Fig.~1c. Thus the corresponding vacuum configurations
have winding number $0$ and $1$.  The detailed variation
of the phase is immaterial since it can always be changed
locally by a gauge transformation. Thus, in Fig.~1a for
example,  as $x$ varies from $-\infty$ to $+\infty$ the field
does not have to stay fixed, but could wander continuously
on the unit circle provided the net change in phase is zero.
However,  the configuration of Fig.~1a  cannot be continuously
deformed  to that of  Fig.~1c  without  leaving the manifold of zero
energy configurations. Therefore, in an evolution between
neighboring
vacua the field configuration must pass over an energy barrier,
as illustrated in Fig.~1b which singles out the configuration
for which $\phi$ vanishes at a point, rendering its phase
there undefined.
Figures 1d-f add the additional perspective
of spatial dependence for the field $\phi(x^1)$.  Figures.~1a-c
can be viewed as projections onto the complex plane orthogonal
to the $x^1$ axis of the curves in Figs.~1d-e.

\begin{figure}
\centerline{
\epsfxsize=120mm
\epsfbox[72 318 540 470]{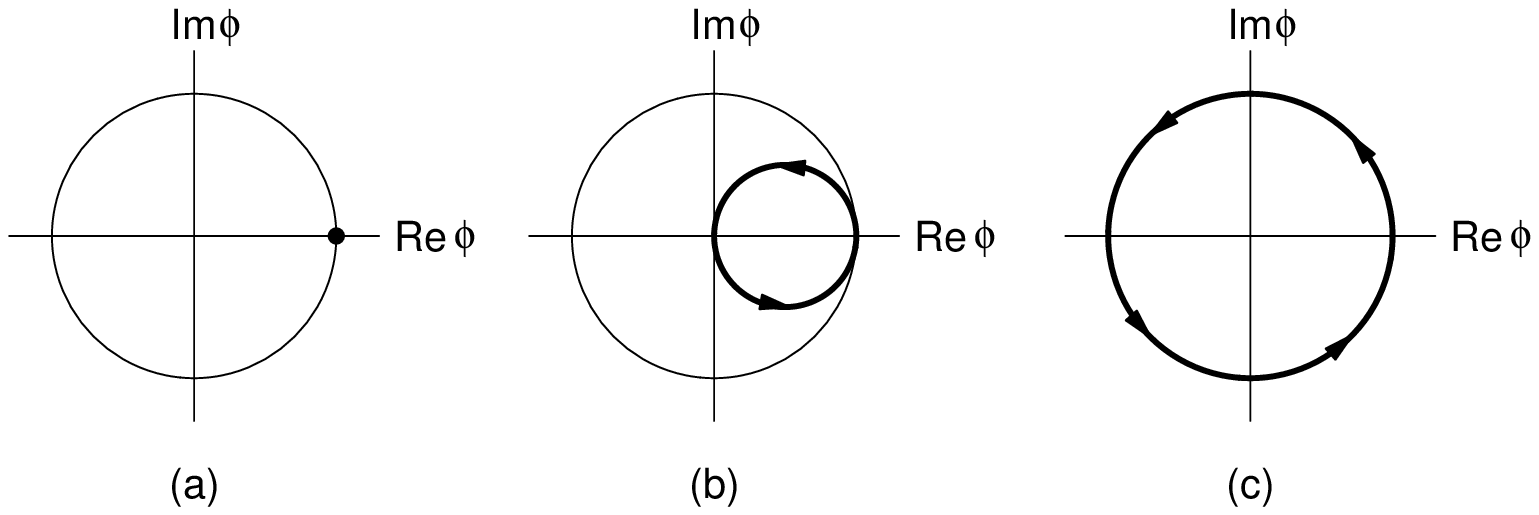}
}
\centerline{
\epsfxsize=135mm
\epsfbox[72 300 540 450]{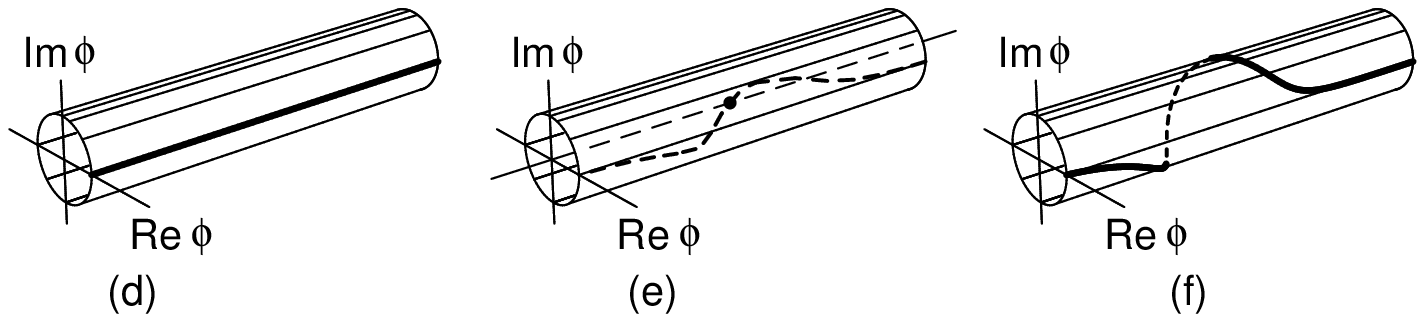}
}
\caption{\tenrm
Example of two inequivalent vacuum configurations (a, c)
and a field configuration at the top of the energy barrier
separating them (b).  Figures a, b and c trace the field
$\phi$ in the
complex plane as the spatial coordinate spans the entire
axis.  A three dimensional perspective has been added
in figures d, e and f  to illustrate the detailed dependence
of $\phi$ on the spatial coordinate.
}
\end{figure}
\vskip4mm

The condition that $\phi$ should take the same value at
$x^1 = \pm \infty$ can be relaxed. Sometimes it is convenient
to use the time independent gauge freedom to make
$\phi(\infty)$ and $\phi(-\infty)$ differ (while keeping both fixed
in time). Thus, the configurations of Figs.~1a-c can be gauge
transformed into the configurations shown in Figs.~2a-c.
In Fig.~2a the phase of $\phi$ changes by $-\pi$ as $x^1$
goes from $-\infty$ to $+\infty$, whereas in Fig.~2c it rotates
by $\pi$.  As in  Fig.~1, the two vacuum
configurations differ by a phase of $2\pi$, i.e. by a unit change
of winding number. In the intermediate configuration (Fig.~2b)
the field takes only imaginary values.  In this gauge the
configuration which minimizes the energy on top of the barrier
(i.e. the sphaleron configuration) takes a very simple form
\begin{equation}
\phi(x^1)=\imath {\rm th}[\sqrt{\lambda} (x^1-c)], \quad A_{\mu}=0.
\label{eq5}
\end{equation}

A possible parameterization for the entire evolution illustrated
in Fig.~2 can be conveniently written
\begin{equation}
\phi(x^1)=\imath {1-\exp [ \imath \tau -2 \sqrt{\lambda} (x^1-c)] \over
1+\exp [\imath \tau -2 \sqrt{\lambda} (x^1-c)] },
\label{eq6}
\end{equation}
\begin{equation}
A_0=0, \quad
A_1={4 \tau \sqrt{\lambda} \over \pi {\rm ch}[2 \sqrt{\lambda} (x^1-c)]}.
\label{eq7}
\end{equation}
As the reader can easily verify, for $\tau=-\pi/2$ and
$\tau=\pi/2$ the field $\phi$ reduces to a number of
unit modulus precisely spanning  the contours of Fig.~2a
and Fig.~2c respectively (as as $x^1$ ranges from $-\infty$
to $+\infty$). The corresponding values of $A_1$ are
chosen  to make the gauge covariant derivative of $\phi$
vanish, thus ensuring vacuum.   We should point out, however,
that Eqs.~\ref{eq6},\ref{eq7} do not represent the solution of
any special equations of motion (Euclidean or Minkowski).
They are merely a  compact  parameterization of interpolating
configurations, in terms of two variables $c$ and $\tau$, which
might be useful in studying sphaleron transitions based on the
method of collective coordinates.

\begin{figure}
\centerline{
\epsfxsize=120mm
\epsfbox[72 318 540 470]{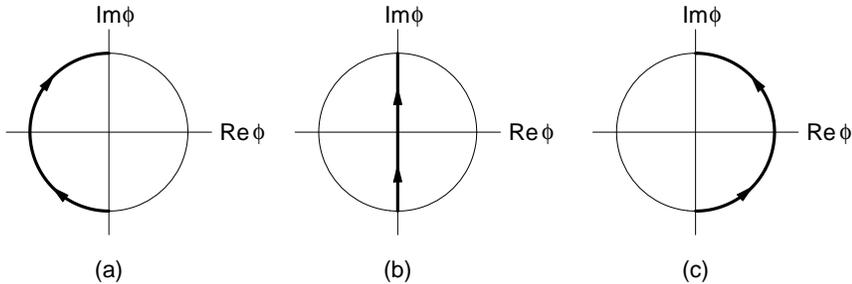}
}
\caption{\tenrm
A different gauge equivalent representation of
the configurations illustrated in Fig.~1.
}
\end{figure}

If one couples chiral fermions to the gauge field in the
2-dimensional Abelian Higgs model the fermionic current
has an anomaly which leads to fermion number violation
in the topology changing processes described above.
Thus this model would appear a very convenient system
for a simplified study of  baryon number violation in high
energy processes.  However, as we will discuss in
the next section, a crucial component of the computational
investigation is the ability to numerically identify  the
normal mode amplitudes of the fields in the asymptotic
regime. No matter how non-linear the system may be at
any given point in its evolution, typically the energy will
eventually disperse and bring the system to a regime where
the fields undergo small oscillations about a vacuum
configuration. This dispersion is expected to occur in any
field  theoretical system, unless prevented by conservation
laws such as those underlying soliton phenomena.
Now, while the 2-dimensional Abelian Higgs model does
not possess soliton solutions, we have observed computationally
that the decay of the sphaleron in this system nevertheless
gives origin to persistent, localized, large oscillations with
an extremely small damping rate (this observation was also
made by Arnold and McLerran in Ref. \cite{am88}).These
oscillations,
illustrated in Fig.~3, make the system quite unwieldy for
a computational investigation of baryon number violation
based on semiclassical techniques.  Thus we eventually
turned our attention to the more realistic 4-dimensional
$SU(2)$ Higgs system, which constitutes the most relevant
component of the full electroweak theory.  Because of the
larger dimensionality of space one would expect the energy
to disperse much more readily in this system, an expectation
borne out by results of Hellmund and Kripfganz\cite{hk91},
who observed the onset of a linear regime following the
sphaleron's decay.

\begin{figure}
\centerline{
\epsfxsize=95mm
\epsfbox[27 305 581 746]{2dsph.fig.eps}
}
\caption{\tenrm
Sphaleron decay in the 2-dimensional Abelian
Higgs model: evolution of the $\phi$ field. The values
of the phase of the complex field are coded by shades
of gray, and the modulus of the field by the height of
the surface. The sphaleron decays rather quickly, but
leaves behind a quasi-stable oscillating remnant.
}
\end{figure}

The 3+1 dimensional  $SU(2)$ Higgs system is defined
in terms of a complex  doublet $\Phi(x)$ and the gauge
potential $A_\mu(x)$ with action
\begin{equation}
S = \int dx ^4 ~ \left\{- {1 \over 4} F_{\mu \nu} F^{\mu \nu} +
(D_{\mu}  \phi)^* D^{\mu} \Phi  - \lambda (|\Phi|^2 -1 )^2
\right\} \ ,
\label{fourAction}
\end{equation}
where the indices run over $0\cdots 3$ and where
\begin{eqnarray}
F_{\mu\nu} &=& \partial_\mu A_\nu -  \partial_\nu A_\mu
 - \imath [A_\mu,A_\nu] \\
D_\mu \Phi &=&  (\partial_\mu - \imath A_\mu) \Phi
\end{eqnarray}
with $A_\mu = A^a_\mu\sigma^a/2$. We use the standard
metric $\eta_{\mu\nu}={\rm diag}(1,-1,-1,-1)$, and  have eliminated
many inessential constants by a suitable choice of units. We
focus on the spherically symmetric configurations of Ratra and
Yaffe\cite{ry88}, which reduce to an effective 2-dimensional
theory. This lower dimensional theory has the full topological
structure of the 4-dimensional system, while having the virtue
of  being computationally manageable.

The spherical {\it Ansatz} is given by expressing the gauge
and Higgs fields  in terms of six real functions  $a_0\, ,\,a_1\, ,
\, \alpha\, , \, \beta\, , \, \mu\ {\rm and}\ \nu\ {\rm of}\ r\
{\rm and}\ t$:
\vfill\break
\begin{eqnarray}
  A_0({\bf x},t) &=& \frac{1}{2 } \, a_0(r,t)
\mbox{\boldmath$\sigma$}\cdot {\bf\hat x}
  \nonumber\\
  A_i({\bf x},t) &=& \frac{1}{2 } \, \big[a_1(r,t)
  \mbox{\boldmath$\sigma$}\cdot {\bf\hat x}
 \hat  x^i+\frac{\alpha(r,t)}{r}(\sigma^i-\mbox{\boldmath$\sigma$}
  \cdot {\bf\hat x}\hat x^i)
  +\frac{1+\beta(r,t)}{r}\epsilon^{ijk}\hat x^j\sigma^k\big]
  \nonumber\\
  \Phi({\bf x},t) &=&   [ \mu(r,t) + i \nu(r,t)\mbox{\boldmath$\sigma$}
\cdot {\bf\hat x} ]
  \xi  \ ,
  \label{SphAn}
  \end{eqnarray}
where ${\bf \hat x}$ is the unit three-vector in the radial direction
and $\xi$ is an arbitrary two-component complex unit vector.
Note that configurations in the spherical  {\it Ansatz} remain in
the spherical  {\it Ansatz} under  gauge transformations of the
form
\begin{eqnarray}
 \label{sphgt}
 A_\mu &&\to A_\mu + \imath U^\dagger \partial_\mu U
 ~~~~~~ \mu=0\cdots3  \\
  \Phi \, &&\to U \Phi \ ,
\end{eqnarray}
where the gauge function  is given by
\begin{equation}
\label{Usph}
U=\exp[\imath\Omega(r,t)\mbox{\boldmath$\sigma$}
\cdot {\bf\hat x}/2] \ .
\end{equation}
Inserting Eqs. \ref{SphAn} directly into Eq. \ref{fourAction},
one obtains an effective 2-dimensional theory with action
  \begin{eqnarray}
  S =  4\pi \int dt\int^\infty_0dr  &&\bigg[-\frac{1}{4}
  r^2f^{\mu\nu}f_{\mu\nu}+(D^\mu \chi)^* D_\mu \chi
  + r^2 D^\mu\phi^* D_\mu\phi
  \nonumber\\
  && -\frac{1}{2 r^2}\left( ~ |\chi |^2-1\right)^2
  -\frac{1}{2}(|\chi|^2+1)|\phi|^2 -  {\rm Re}(i \chi^* \phi^2)
  \nonumber \\
  && -\lambda  \, r^2 \, \left(|\phi|^2 - 1\right)^2 ~
  \bigg] \ ,
  \label{effAction}
  \end{eqnarray}
where the indices  now run from $0$ to $1$ and in contrast
to Ref.  \cite{ry88} are raised and lowered with
$\eta_{\mu\nu}={\rm diag}(1,-1)$, and where
\begin{eqnarray}
  f_{\mu\nu}&=& \partial_\mu a_\nu-\partial_\nu a_\mu\
\label{defConva}   \\
  \chi &=&\alpha+\imath \beta
\label{defConvb} \\
  \phi &=& \mu+\imath \nu\
\label{defConvc} \\
  D_\mu \chi &=& (\partial_\mu- \imath   \, a_\mu)\chi
\label{defConvd}  \\
  D_\mu \phi&=& (\partial_\mu - \frac{\imath}{2}  \, a_\mu)\phi\ .
\label{defConve}
\end{eqnarray}
The reduced action, Eq. \ref{effAction}, is invariant under
the gauge transformation
\begin{eqnarray}
 a_\mu &&\to a_\mu + \partial_\mu \Omega
\label{amueq}\\
  \chi \, &&\to e^{\imath \Omega} \chi \ ,
\label{chieq}\\
  \phi \, &&\to e^{\imath \Omega/2} \phi
\label{phieq}
\end{eqnarray}
which corresponds to the residual $U(1)$ gauge invariance
of  Eqs. \ref{sphgt}-\ref{Usph}.
{}From Eqs. \ref{effAction}-\ref{chieq}, we see that the
spherical  {\it Ansatz} effectively
yields  a system very similar to the Abelian Higgs model
considered above.  In this reduced system the variables
$a_0(r,t)$ and $a_1(r,t)$ play the role of the gauge field,
whereas the variables $\chi(r,t)$, which parameterizes
the residual components of the 4-dimensional gauge field,
and $\phi(r,t)$, which parameterizes the 4-dimensional Higgs
field, both behave as 2-dimensional Higgs fields.  Of course,
the presence of metric factors (powers of $r$) in the
action Eq.~\ref{effAction} is a reminder  that we are really
dealing with a 4-dimensional system.

Regularity of the 4-dimensional field configuration for $r=0$
requires
\begin{eqnarray}
\label{atzero}
&&\chi(r=0)=-\imath \\
\nonumber
&& {\rm Im}\phi(r=0)=0  \ .
\end{eqnarray}
Although one could modify these boundary conditions
by a singular gauge transformation, this would introduce
unnecessary complications.  So we will impose
Eqs.~\ref{atzero}, in addition to some other boundary
conditions  that also follow from the regularity of the
4-dimensional configuration.
It is also worth noting here that non-singular
gauge transformations satisfy the condition that the gauge
function $\Omega(r,t)$ vanishes at $r=0$ (or is a multiple
of $2\pi$).

We shall work in the $a_0=0$ (or $A_0=0$) gauge throughout. In
the  overlaying 4-dimensional theory, if one compactifies 3-space
to $S^3$ by identifying the points at infinity, it is well known that
the vacua correspond to the  topologically inequivalent
ways that $S^3$ can be mapped into $SU(2)\sim S^3$\cite{JR}.
These maps are characterized by the third homotopy group of
$SU(2)$, and a vacuum can be labeled by an integer called the
homotopy index or winding number. The effective 2-dimensional
theory inherits a  corresponding vacuum structure. From
Eq.~\ref{effAction} it is  apparent that the vacuum states are
characterized by $|\chi|=|\phi|=1$,  with the additional constraint
that $\imath \chi^* \phi^2 =-1$ (as well as $D_1 \chi =D_1 \phi =0$).
A convenient zero-winding vacuum is given by
$\chi_{vac}=-\imath$, $\phi_{vac}=1$. Nontrivial vacua can be
obtained from this vacuum via  gauge transformations in which
the gauge function $\Omega  \to 2\pi n$ (for non-zero integers
$n$) as $r\to\infty$. Note that  the compactification of 3-space
ensures that at infinity $\Omega$ is a multiple of $2\pi$. Since
$\Omega$ has been locked down to zero at
$r=0$, the winding of such a gauge transformed configuration
is just the integer $n$.  For example, a typical winding-number
one vacuum obtained from the previous trivial vacuum is given
by $\chi_{vac}=-\imath \exp[\imath \theta(r)]$, $\phi_{vac}=
\exp[\imath \theta(r)/2]$ and $a_{1 \, vac}=\partial_r \theta(r)$,
where $\theta(r)$ varies from $0$ to $2\pi$ as $r$ changes from
$0$ to $\infty$. By taking advantage of the freedom of performing
a time independent gauge transformation, however, one can
also choose a gauge where $\chi(r)$ and $\phi(r)$ tend to values
different from $-\imath$ and $1$ as $r\to\infty$ (the condition
$\imath \chi^*\phi^2 \to -1$ must be preserved). Indeed, it will
be convenient to choose one such gauge to parameterize the
sphaleron.

In making a topological transition between two inequivalent
vacua, one must leave the manifold of vacuum configurations
and pass over an energy barrier. Along such a trajectory there
will be a configuration of maximum energy. Of all these maximal
energy configurations, the sphaleron has the lowest
energy and represents a saddle point along the energy
ridge separating inequivalent vacua\cite{KM}. The sphaleron
can be expressed in the spherical {\it Ansatz}, and it is
convenient to choose a gauge in which $a_\mu=0$ and
\begin{eqnarray}
\label{sphSphal}
&&\chi_{sph}(r)=\imath [2 f(r)-1] \\
\nonumber
&& \phi_{sph}(r)=\imath h(r)  \ ,
\end{eqnarray}
where $f$ and $h$ vary between $0$ and $1$ as $r$
changes from $0$ to $\infty$ and are  chosen to minimize
the energy functional.

\begin{figure}
\centerline{
\epsfxsize=120mm
\epsfbox[72 318 540 470]{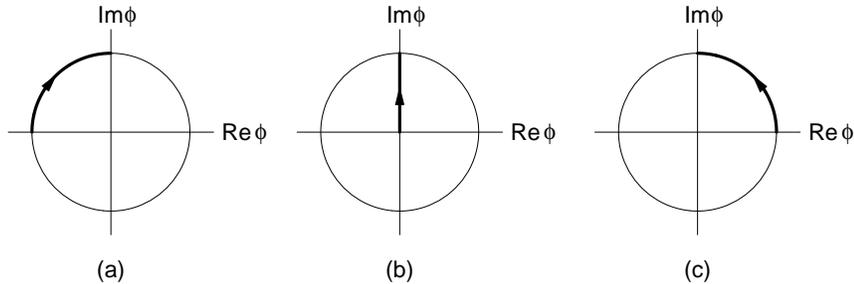}
}
\caption{\tenrm
The behavior of the $\phi$ field for a typical topological
transition. The $\chi$ field has a  behavior similar to the
one in Fig. 2.
}
\end{figure}

This choice of gauge for the sphaleron is slightly
peculiar in the following sense. Finite energy
configurations, like Eq.  \ref{sphSphal}, asymptote
to pure gauge at spatial infinity. Typically a gauge
is chosen so that the appropriate gauge function
is unity at spatial infinity, and then space can be
compactified to the 3-sphere. But consider the spherical
gauge function Eq. \ref{Usph} with $\Omega(r)$ independent
of time and varying from $0$ to $\pi$ as $r$ goes from
$0$ to $\infty$. From Eqs. \ref{chieq}-\ref{phieq} we
see that the
corresponding $\chi,\phi \to \imath$ for large $r$,
which is the same as the sphaleron boundary conditions
at spatial infinity. Note, however, that  $U\mid_{r\to\infty}
=\imath
\mbox{\boldmath$\sigma$}\cdot {\bf\hat x}$
is direction dependent, so space cannot be
compactified.  An arbitrary element of $SU(2)$ can be
parameterized by  ${\rm b}_0 {\bf 1} + \imath \mbox{\boldmath$\sigma$}\cdot
{\bf b}$ where
${\bf 1}$ is the two by two unit matrix and
${\rm b}_0^2+{\bf b}^2=1$. Hence $SU(2)\sim S^3$,
and defining the north and south poles by
$\pm {\bf 1}$, we see that  $\imath
 \mbox{\boldmath$\sigma$}\cdot {\bf b}$
with ${\bf b}^2=1$ parameterizes the equatorial
sphere. Thus, the sphere at infinity is
mapped onto the equatorial sphere of $SU(2)$.
In this gauge, a topology
changing transition proceeding over the sphaleron
corresponds to a transition where the fields wind
over the lower hemisphere of $SU(2)$ before the
transition and over the upper hemisphere after the
transition, with a net change in winding number still
equal to one. In this gauge, the behavior of the $\chi$
field in a
topological transition will be very similar to the behavior
of the Higgs field in the 2-dimensional model, already
illustrated in Fig.~2. The behavior of the $\phi$ field
is illustrated in Fig. 4. We could of course work in a
compactified gauge  where a topological transition
would occur between a field with no winding and a field
with unit winding, as in Fig.~1, but the sphaleron would
look more complicated. The advantage of  Eq. \ref{sphSphal}
from a computational perspective is that perturbations about
the sphaleron can be more easily parameterized.

In the sphaleron configuration  the $\chi$ field has a zero at
some non-zero value of $r$, whereas the $\phi$ field has a zero
for $r=0$ corresponding to the vanishing at the origin of the
actual 4-dimensional Higgs field. The zero of $\chi$ is
reminiscent of the zero which characterizes the sphaleron
of the Abelian Higgs model.  However, as shown in
Ref. ~\cite{fggrs},
it is the zero of the true Higgs field (i.e. the zero
of $\phi$) which carries a deeper significance and should be
associated with the actual occurrence of the topological
transition.  Nevertheless, since the phase changes of $\chi$ are
more dramatic than those of $\phi$, for purposes of
illustration it is often more convenient to plot $\chi$.
Figure~5 illustrates a typical topological transition. The
configuration starts as  a small excitation about the
trivial vacuum defined above, passes over the sphaleron
and then emerges as a configuration that undergoes a
$2\pi$ phase change.

\begin{figure}
\centerline{
\epsfxsize=80mm
\epsfbox[27 240 581 746]{sph9.2_3.fig.eps}
}
\caption{\tenrm
Behavior of the $\chi$ field in a topology changing transition.
The various shades of gray code the phase of the complex
field. The field starts as an excitation about the trivial vacuum,
passes over the sphaleron and then emerges as an excitation
about the vacuum of unit winding. Note the persistent strip of
$2\pi$ phase change after the wave bounces off the origin.
}
\end{figure}

\section{Numerical study of classically allowed processes}
\hspace{1cm}
For energy larger than the sphaleron energy $E_{sph}$, i.e. for
$\epsilon > \epsilon_{sph}= g^2 E_{sph}$, classical evolution
which changes the topology of the fields becomes possible.
Solutions to the classical equations of motion provide the dominant
contributions in a weak coupling expansion of the path integrals
describing transition processes between coherent states. The
existence of topology-changing solutions indicates that the
corresponding processes, which because of the change of topology
violate baryon number, are not suppressed.  However it would be
premature to conclude that baryon number violation can occur with
non-negligible amplitude in high energy collisions. Indeed, the number
of particles in the initial state of such processes is small and, in terms
of the rescaled fields used in the description of the classical evolution,
this converts into a rescaled particle number $\nu = g^2 N$ which
tends to zero as $g \to 0$. Thus, in order to establish evidence for
baryon number non-conservation in high energy collisions, one
must show that topology changing classical evolution can occur
with arbitrarily small $\nu$ in the initial state.

The primary impediment for rapid baryon number violation seems
to be a phase space mismatch between  initial states of low multiplicity
and final states of many particles. The authors of Ref. \cite{gr} look at
simplified models and observe that,  classically, it is difficult to transfer
energy from a small number of hard modes to a large number of soft
modes. However,  Ref. \cite{wdymt} finds that for pure Yang-Mills
theory in 2-dimensions the momenta can be dramatically redistributed,
but unfortunately the initial particle number seems to be rather
large in their domain of applicability. It is the purpose of our
investigation to shed light on the situation in 4-dimensions when the
Higgs field is added.

Given any classical evolution, because of the dispersion of the
energy, the fields will asymptotically approach vacuum values.
Thus, for times $t < -T_i$ and $t > T_f$ and sufficiently large
$T_i, \, T_f$, the equations of motion will reduce to linearized
equations describing the small oscillations of the system about
a vacuum configuration.  In this linear regime, the evolution of
the fields will be given by a superposition of independent harmonic
oscillators (the normal modes). In terms of the frequencies $\omega_n$
and amplitudes $a_n$ of these oscillators the (rescaled) energy and
particle number are given by
\begin{equation}
\epsilon = \sum_n \omega_n a_n^* a_n
\label{eq31}
\end{equation}
and
\begin{equation}
\nu = \sum_n  a_n^* a_n .
\label{eq32}
\end{equation}
Thus for any classical evolution the energy $\epsilon$ and the
particle numbers $\nu_i$ and $\nu_f$ of the asymptotic initial and
final states are well defined.  (The energy is of course conserved
and well defined even in the non-linear regime.)  In addition,
because of the fact that the fields approach vacuum values for
$t \to \pm \infty$, the winding numbers of initial and final
configuration are also well defined. Because of the sphaleron
barrier, the energy $\epsilon$ of all the classical solutions with
a net change of winding number is bounded below by the sphaleron
energy $\epsilon_{sph}$.  The problem one would like to solve, then,
is whether the initial particle number $\nu_i$ of these solutions can
be arbitrarily small, or more generally, one would like to map the
region spanned by all possible values of $\epsilon$ and $\nu_i$ for
the topology changing classical evolution. The highly non-linear
nature of the equations of motion makes an analytic solution
unlikely, even if one is willing to make the crudest approximations.
The problem is however amenable to solution by computational
techniques. In this section we will illustrate the strategy we have
followed to formulate it on the computer and the progress we have
been able to make towards its solution.

The fundamental computational ingredient consists in the
implementation of a numerical solution of the equations of
motion.  We start from the Hamiltonian formulation of the
equations of motion for the continuum system in the $a_0=0$
gauge.   In such a formulation the variables
\begin{equation}
a_1(r), \quad \chi(r), \quad \phi(r)
\label{eq33}
\end{equation}
form a set of canonical coordinates, conjugate to the momenta
\begin{eqnarray}
E(r)= &&r^2 \partial_0 a_1, \nonumber \\
\pi_{\chi}(r)=&&\partial_0 \chi, \nonumber \\
\pi_{\phi}(r)=&& r^2\partial_0 \phi \ .
\label{eq34}
\end{eqnarray}
The evolution of these variables is generated by the Hamiltonian
  \begin{eqnarray}
  H =  \int^\infty_0dr \, && \bigg[\frac{E^2}{2r^2} +|\pi_{\chi}|^2
  +{|\pi_{\phi}|^2 \over r^2} + |D_r \chi|^2  + r^2 |D_r\phi|^2
  \nonumber\\
  && +\frac{1}{2 r^2}\left( ~ |\chi |^2-1\right)^2
  +\frac{1}{2}(|\chi|^2+1)|\phi|^2 + {\rm Re}(i \chi^* \phi^2)
  \nonumber \\
  && +\lambda  \, r^2 \, \left(|\phi|^2 - 1\right)^2 ~
  \bigg] \ .
  \label{eq35}
  \end{eqnarray}
Gauss' law
\begin{equation}
\partial_r E =\imath(\pi_{\chi}^* \chi - \chi^* \pi_{\chi})
+\imath(\pi_{\phi}^* \phi - \phi^* \pi_{\phi})
\label{eq36}
\end{equation}
expresses the residual invariance of the system under time
independent local gauge transformations and is imposed as
a condition on the initial state.  It is then automatically conserved
by the equations of motion.

To solve the equations of motion numerically the system must be
discretized.  It is convenient to use the formalism of lattice gauge
theories.  The $r$-axis is subdivided into $N$ equal subintervals
of length $\Delta r$ with finite length $L= N \, \Delta r$.  Thus,
the lattice sites have spatial coordinates $r_i=i\Delta r$ with
$i=0 \cdots N$, and the
midpoints between lattice sites have coordinates
$r_{i+1/2}=(i+1/2)\Delta r$ with $i=0 \cdots N-1$. The fields
$\chi$, $\phi$, $\pi_\chi$ and $\pi_\phi$ are then represented
by discrete variables defined over the end points of the intervals,
$\chi_i=\chi(r_i)$, $\phi_i=\phi(r_i)$, etc.; whereas the fields
$a_1$ and $E$ are defined over the intervals themselves
by $a_{1 i}\equiv a_i=a_1(r_{i+1/2})$ and  $E_i=E(r_{i+1/2})$
(for notational
simplicity we have dropped the spatial subscript on the
discretized gauge field). The covariant derivatives are then
replaced by covariant finite differences, e.g.
\begin{equation}
D_r \chi \to { \exp[- \imath a_i\, \Delta r] \,
\chi_{i+1}- \chi_i \over \ \Delta r}
{}~~~~~~~ i=0 \cdots N-1\ ,
\label{eq37}
\end{equation}
and like the gauge fields they are to be thought of as living
on the links between lattice sites.
The rest of the discretization is straightforward.  One obtains
a discretized Hamiltonian $H_D$ expressed in terms of a finite set of
variables, which still possesses exact local gauge invariance
under the transformations of Eqs.~\ref{amueq}-\ref{chieq}
provided that $a_1 \to a_1 + \partial_r \Omega$ is replaced
with the finite difference formula
\begin{equation}
a_i \to a_i +\frac{\Omega_{i+1} - \Omega_i} {\Delta r}
 ~~~~~~~ i=0 \cdots N-1 \ ,
\end{equation}
where $\Omega_i=\Omega(r_i)$. From $H_D$ one can
easily obtain the canonical evolution equations for the
discretized variables.  Gauss' law, which now takes a
discretized form, must be imposed on the initial state and
is then preserved (exactly) by the time evolution because
of the gauge invariance of the discretized system.  In practice
we have used values of $N$ equal to 256, 512  and 1024
and values of $\Delta r$ equal to 0.2, 0.1 and 0.05 in
respectively study the properties of the
system (with $\lambda=0.1$).  We found these parameters
to be adequate for obtaining, on the one hand,  a reasonable
approximation  to the continuum system and,  on the other,
a cut-off on $r$ sufficiently large to allow for an effective
linearization of the equations of motion before the waves
hit the boundary.  The restriction to uniform spacing of the
subintervals on the $r$-axis is not fundamental and we have
also implemented a discretization where $\Delta r$ increases
as one moves out on the $r$-axis.  In this manner one can
effectively make the system larger and delay the effects of
the impact of the waves
with the boundary without worsening the spatial resolution near
$r=0$, where most of the non-linear dynamics takes place.  We
have found however that the advantages one gains hardly warrant
the additional complications introduced by the non-uniform spacing.

For the numerical integration of the time evolution we have used the
leap-frog algorithm.  Since this algorithm, or the equivalent velocity
Verlet algorithm, constitutes one of the fundamental techniques for
the integration of ordinary differential equations of the Hamiltonian
type and as such is textbook material, we will not discuss it in any
great detail.  Essentially, given conjugate canonical variables
$q_i$ and $p_i$ which obey equations
\begin{eqnarray}
{dq_i \over dt} = g_i(p), \nonumber\\
{dp_i \over dt} = f_i(q),
\label{eq38}
\end{eqnarray}
one evolves the values of $q$ and $p$ from some initial $t$ to
$t+\Delta t$ as follows. In a first step $p_i$ is evolved to the
mid-point of the time interval by
\begin{eqnarray}
p_i \to p_i'=&& p_i+f_i(q) {\Delta t \over 2}, \nonumber \\
q_i \to q_i'=&& q_i \ .
\label{eq39}
\end{eqnarray}
(Although $q_i$ is left unchanged, it is convenient to consider the
step formally as a transformation of the entire set of canonical
variables.) In a second step one evolves the coordinates from
their initial value $q_i=q_i'$ to their value at the end of the interval
\begin{eqnarray}
p_i' \to p_i''=&& p_i',\nonumber \\
q_i' \to q_i''=&& q_i' +g_i(p') \Delta t \ .
\label{eq40}
\end{eqnarray}
Finally the momenta are evolved from their value at the midpoint
to the final value
\begin{eqnarray}
p_i'' \to p_i'''=&& p_i''+f_i(q'') {\Delta t \over 2}, \nonumber \\
q_i'' \to q_i'''=&& q_i'' \ .
\label{eq41}
\end{eqnarray}

One can easily verify that these equations reproduce the correct
continuum evolution from $t$ to $t+\Delta t$ up to errors of order
$(\Delta t)^3$.  Moreover, the algorithm has the very nice property
that all  three steps above constitute canonical transformations
and that it is reversible (in the sense that starting from $q_i'''$,
$-p_i'''$, up to numerical errors one would end up exactly with
$q_i$, $-p_i$).  Another very nice feature of the algorithm is
that, although the evolution of the variables is affected by
errors of order $(\Delta t)^3$, the energy of a harmonic oscillator,
and therefore also of any system which can
be decomposed into a linear superposition of harmonic oscillators,
is conserved exactly (always up to numerical errors).
In a sequence of several iterations of the algorithm after the
momenta have been been evolved by the initial $\Delta t/2$,
the first and third step can be combined into a single step whereby
the momenta are evolved from the midpoint of one interval to the
midpoint of the next one, ``hopping over'' the coordinates, which
are evolved from endpoint to endpoint.  This motivates the name
assigned to the algorithm.

With a good grasp on the numerical solution of the equations of
motion, we can now turn to the second crucial component of the
computation, namely the identification of the particle number in
the initial state. One could easily parameterize an initial
configuration of the system consisting of incoming waves in the
linear regime. However, it would be extremely difficult to adjust
the parameters so as to insure that a change of winding number
occurs in the course of the subsequent evolution.  For this reason
it is much better to parameterize the initial configuration of the
system at the moment when a change of topology occurs. Thus our
strategy consists in implementing a time-reversed solution of the
equations of motion, where the initial configuration is the
configuration of the system at the moment when it passes over the
sphaleron barrier and the asymptotic configuration for large $t$ will
be interpreted as a time-reversed incoming state.

Topology changing transitions within the spherical ansatz are
characterized by the vanishing of
$\phi$ at $r=0$ and the the vanishing of $\chi$ at  nonzero $r$.
For a sequence of configurations that pass directly through the
sphaleron these two zeros occur at the same time. However,
this is not the most general case and the zeros of $\phi$ and
$\chi$ need not occur simultaneously \cite{gi}. So we have
parameterized initial configurations in terms of coefficients
$c_n$ of some suitable expansion of the fields and their conjugate
momenta, constrained only by the fact that the field $\chi$ must
have a zero at some finite $r$, as in the sphaleron configuration.
Furthermore, we can use gauge invariance to make this field pure
imaginary. The field $\phi$ is only restricted to obey the boundary
conditions and does not necessarily vanish at the origin (although
it will vanish at some point in its evolution). \footnote{Of course we
could equally well arrange $\phi$ to be pure imaginary and to
vanish at the origin, with no restriction $\chi$ on  other than its
boundary conditions.}

In practice, computational considerations will limit
the number of parameters we will be able to use.
We will then evolve the system until the dispersion of the energy
brings it to the linear domain.  At this point, we can calculate
the amplitudes of the normal modes of oscillation and the particle
number of the system in its asymptotic state.  It is clear that
every set of coefficients $c_n$ will determine one definite value
for the energy $\epsilon$ and particle number $\nu$ of the asymptotic state.
In calculating the particle number, one
should  use only the lower lying modes since higher modes
probe wave lengths of order the lattice spacing. For our
lattice parameters, we found that considering the first 50 to
100 modes is reasonable.
We should point out that an arbitrary initial configuration is
not necessarily guaranteed to change topology. However, by
evolving the configuration both forward and backward in time we
can easily verify whether  topology changes, and initial
configurations that do not change the topology can be rejected.
By reversing the time evolution, then, we will have defined
an initial asymptotic configuration with energy $\epsilon$ and
particle number $\nu$, which in the course of its evolution
undergoes a change of topology.  By varying the values of the
parameters $c_n$ we will be able to study the properties of such
field evolution and, in particular, explore the domain of permissible
values for $\epsilon$ and $\nu$.  It should be obvious at this
point that the determination of the normal mode amplitudes is
another crucial ingredient of the computation.

The normal modes of oscillation can be calculated starting from
the linearized equations of motion.   These equations can be obtained
from an expansion of the Hamiltonian up to second order in the
deviation of fields from a vacuum configuration.  For conciseness
of presentation we will not reproduce here their explicit form, but
will limit ourselves to a discussion of the general properties of the
normal modes.  The normal modes can be obtained by assuming
an oscillatory evolution for the fields of the type
\begin{eqnarray}
\delta \chi(r,t) = && \chi^{(n)}(r) \sin[\omega_n t]
\nonumber \\
\pi_{\chi}(r,t) = && \pi_{\chi}^{(n)}(r) \cos[\omega_n t] \ .
\label{eq42}
\end{eqnarray}
In this equation we have made reference only to one pair of
conjugate fields, but the equation should be complemented with a similar
assumption for the pairs $a_1$ and $E$,
$\delta \phi$ and $\pi_\phi$,
which are a priori all coupled together with $\delta \chi$ and
$\pi_\chi$
in the equations of evolution. We have used the symbol $\delta$
to denote the deviations of the fields $\chi$ and $\phi$ from
their vacuum values $\chi= -\imath$ and $\phi=1$.  All of the other
fields vanish in the trivial vacuum configuration.
For the discretized system, $r$ should be replaced by an index $i$.

Substituting the {\it Ansatz} of Eqs.~\ref{eq42} into the equations of motion
one obtains a set of eigenvalue equations.  Their solution determines
the possible values of $\omega_n$ as well as the corresponding
eigenfunctions (or more properly eigenvectors in the discretized case)
$\delta \chi^{(n)} (r)$, $\pi_{\chi}^{(n)} (r)$ etc.
We have determined eigenvalues and eigenfunctions both numerically,
for the discretized system, and analytically, for the continuum
system.  One finds that the normal modes of oscillations are naturally
grouped together into four sets of modes:

i) a set of modes where only the imaginary part of the fields $\delta \chi$
and $\pi_{\chi}$ is non-vanishing.  These correspond to an oscillation of the
modulus of the $\chi$ field .

ii) a set of modes where only the real part of the fields $\delta \phi$
and $\pi_{\phi}$ is non-vanishing.  These correspond to an oscillation of the
modulus of the $\phi$ field .

iii) and iv) two sets of modes where the real part of  $\delta \chi$,
$\pi_{\chi}$, the imaginary part of $\delta \phi$, $\pi_{\phi}$ as well
as $a_1$ and $E$ are coupled together and non-vanishing.  These
correspond to oscillations of the phases of  $\chi$ and $\phi$
(in coherence or opposition of phase), accompanied corresponding
oscillations of the gauge fields.

Given the expressions for the eigenfunctions it is possible
to extract the amplitudes of the normal modes of oscillation
$a_n(t)$ by taking suitable convolutions of the eigenfunctions
with the fields and momenta. This procedure exploits various
properties of orthogonality which the eigenfunctions satisfy.
One subtle point, however, involves the need to fix the gauge.
The normal modes are obtained on the basis of an expansion
into small oscillations around the trivial vacuum configuration
with constant fields (one could of course expand around any
fixed vacuum configuration, but the formulae are much simpler
if one expands around constant fields).  However there is no
guarantee that the evolution of the fields will lead to an
asymptotic regime of small fluctuations precisely around
such a vacuum configuration.  Indeed, in general this will
not happen and the actual configuration  will typically differ
from the one used to derive the normal modes by a large
gauge transformation. The remedy is easy enough.  One
can perform a gauge transformation to a fixed gauge which
differs at most by small fluctuations from the one where the
expansion has been performed.  In our computations we have
used the unitary gauge defined by ${\rm Arg}\chi(r) = -\pi/2$,
transforming the fields to this gauge at some definite time
$t_0$, at which point the appropriate amplitudes may be
extracted. At subsequent times the fields might no
longer be small perturbations about the trivial vacuum,
so every time the amplitudes are calculated we first enter
the above gauge. An alternative approach, which we have
also implemented, consists in deriving the linearized equations
of motion for a complete set of gauge invariant quantities
\cite{gi}(we have used  $ |\chi| -1 $, $|\phi| -1 $, the electric
field $E$, the difference of phases ${\rm Arg}\chi -
2{\rm Arg}\phi$ and the time derivatives of all these quantities).
Following a procedure similar to the one  outlined above
(cf. Eq.~\ref{eq42}) one can derive the  eigenvalues and
eigenfunctions for the small oscillations of these quantities
(the eigenvalues are identical, of course, to those obtained
using gauge variant quantities) and extract the amplitudes
through suitable convolutions with the evolving fields. We
have verified that the two procedures produce identical
results.

\begin{figure}
\centerline{
\epsfxsize=80mm
\epsfbox[27 240 581 746]{4dsph.fig.eps}
}
\caption{\tenrm
Sphaleron decay in the four dimensional $SU(2)$ Higgs model:
evolution of the $\chi$ field. The values of the phase of the
complex field are coded by different shades of gray.  When
the evolution reaches the linear regime, a gauge transformation,
indicated by the sudden change of shading, is performed to
extract the normal mode amplitudes.}
\end{figure}

We can now turn our attention to the figures that illustrate
our results for a system with coupling constant $\lambda = 0.1$.
Figure 6 illustrates the behavior of the field $\chi$  following
the decay of the sphaleron after a slight initial perturbation.
We have found it very convenient and informative to use color
to code the phase of the complex fields. Unfortunately the illustrations in
these pages cannot be reproduced in color
and we have tried to render the variation of the phase with
a gray scale.  From Fig.~6 it is clear that the energy, which
is concentrated in the neighborhood of $r=0$ in the sphaleron
configuration, disperses and gives rise to a pattern of outgoing
waves.  The sudden variation of tonality at some point in the time
evolution indicates the change of phase induced by the gauge
transformation to the unitary gauge.  In Fig.~7 we display the
behavior of the particle number in the four normal modes of
oscillations as function of time.  It is apparent that, after an
initial transition period where the particle number is not well
defined, the quantities settle to values which are reasonably
constant in time.  We take this as evidence that the system has
reached an asymptotic regime where one can meaniningfully
define a conserved particle number. Finally, the evolution of
the system can be time reversed, as we have discussed above,
and the (time-reversed) final configuration can be considered
as an asymptotic initial configuration with definite energy and
particle number.  The previously shown Fig.~5 illustrates the
evolution obtained taking one such asymptotic initial
configuration.  The incoming waves are seen to merge in the
neighborhood of the origin, where a change of topology takes
place.  The fact that the winding number of the field configuration
has changed is indicated by the strip of rapidly varying tonality
which persists in the neighborhood of the origin and codes the
variation of the phase of $\chi$.  With color, this strip would
appear as a vivid rainbow, left over as a marker of the change
of topology of the evolving fields.

\begin{figure}
\centerline{
\epsfxsize=100mm
\epsfbox{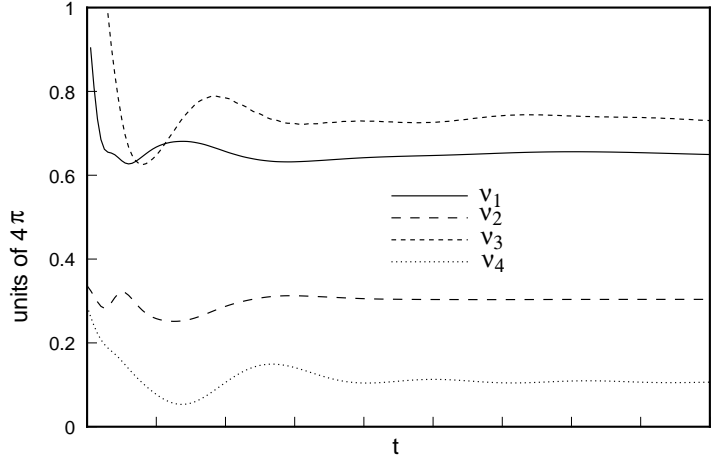}
}
\caption{\tenrm
Sphaleron decay in the four dimensional $SU(2)$ Higgs model:
behavior of the particle number in the four normal modes of
oscillation of the linearized system as function of time. The sum
in Eq. 27 extends over the first 50 modes for each of the four
mode types.
}
\end{figure}

The configuration which sits on top of the sphaleron barrier
can be parameterized by expanding into suitable components
a complete set of independent fields.  We have chosen these
fields to be the perturbations $ \Delta \chi$ and $\Delta \phi$
(not necessarily small)  of the $\chi$, $\phi$ fields of the sphaleron,
the field $a_1$ and the momenta $\pi_{\chi}$, $\pi_{\phi}$. We
avoid a redundant gauge degree of freedom by taking $\Delta
\chi$ purely imaginary (like the $\chi$ field of the sphaleron
itself).  The final field needed to specify the initial configuration,
i.e. the electric field $E$, can then be derived from Gauss law.
We have parameterized these fields in terms of Bessel functions,
chosen in such a way to respect appropriate boundary conditions
at $r=0$ and $r=L$.  The coefficients $c_n$ of the expansion can
now be varied and, in this way, one can explore the region in the
$\epsilon$-$\nu$ plane spanned by all of the topology changing
classical solutions.

Of course, the space of  topology changing configurations is
infinite and a random exploration of such space would not lead
to very useful results. We must keep in mind that the interesting
question is whether there is a lower bound in $\nu$ or, more
generally, what is the lower boundary of the region spanned
by all topology changing solutions in the  $\epsilon$-$\nu$
plane.  This question can be investigated by methods of
stochastic sampling.  One can perform random small steps in
the space of all configurations by varying the parameters $c_n$
stochastically.  After each change the new configuration is
evolved until one can extract the particle number in the linear
regime. Then the variations of energy and particle number
$\Delta \epsilon$ and $\Delta \nu$ induced by the change
$\Delta c_n$ become well defined. A standard Metropolis
Monte Carlo sampling technique consists in accepting or
rejecting the change according to the value of the quantity
$\Delta F=\beta \Delta \epsilon + \mu \Delta \nu$, where $\beta$
and $\mu$ are parameters that weight what region of the
$\epsilon$-$\nu$ plane is explored. To be more precise,  the
change is accepted with conditional probability $ P = {\rm Min}
[1, \exp (- \Delta F)] $, which  has the effect of producing
configurations distributed according to a measure
proportional to $\exp (-\beta  \epsilon -\mu  \nu)$. The
parameters $\beta$ and $\mu$ play the role of inverse
temperature and chemical potential. By choosing these
parameters appropriately one can drive the sampling towards
low values of energy and particle number and thus explore
the interesting region in the $\epsilon$-$\nu$ plane.  We have
begun implementing this procedure and Fig.~8 illustrates our
first results. It is interesting to note that the decay of a
(slightly perturbed) sphaleron gives rise to a particle number
$\nu \approx 1.9 (4\pi)$.  For $g=0.6$ this corresponds to
$N\approx 66$ physical particles. From Fig.~8 one can
see that our sampling procedure has produced configurations
with comparable energy and much smaller particle number.
Of course the ultraviolet cutoff induced by the lattice discretization
puts a lower limit on the ratio $\nu/\epsilon$, which occurs when
only the highest mode is excited. We have used lattice parameters
$\Delta r = 0.2$, $N=256$ with $L=51.2$ and have considered
only the first 50 modes for each of the four types of normal modes.
Hence  the minimum value of $\nu/\epsilon$ is of order
$ 1/\omega_{max}
\sim  L/n_{max}\pi \sim 0.3$. Given our lattice resolution, we have
saturated the lowest bound in particle number that we are
sensitive to. This may be an indication that there is no lower
bound on $\nu$, but our calculation is still at a very preliminary
stage and much more work will be needed to establish reliable
results.

\begin{figure}
\centerline{
\epsfxsize=100mm
\epsfbox{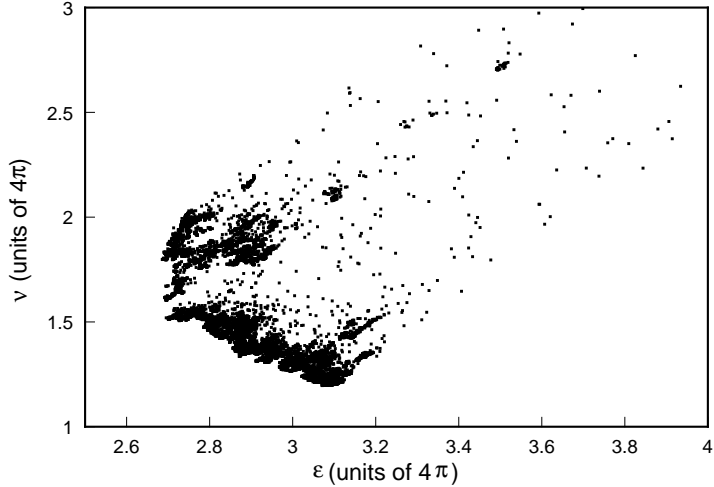}
}
\caption{\tenrm
Monte Carlo results with lattice parameters of $\Delta r = 0.2$,
$N=256$ and L=51.2 and Higgs coupling $\lambda=0.1$.
}
\end{figure}

\section{High energy baryon number violating processes below
the sphaleron barrier}
\hspace{1cm}
In the work of Ref.~\cite{rst}
Rubakov, Son and Tinyakov relate the probability
of a topology changing process from
a coherent state
to the action of a complexified classical evolution along
a special contour in the complex time plane.
As shown in Fig.~9, this contour consists of a
semiaxis parallel to the real axis
$\alpha = (-\infty +\imath T \longrightarrow \imath T)$,
followed by a segment
$\beta=(\imath T  \longrightarrow 0$) of length $T$ along the imaginary axis,
followed by the real positive semiaxis
$\gamma = (0 \longrightarrow +\infty)$.
The word ``complexified classical evolution'' refers to the fact that
the equations of motion must be analytically continued to complex fields.
This has to be handled with some care for the case in which fields,
such as $\chi(r)$ and $\phi(r)$,  are already complex to begin with.
In this case $\chi(r)$ and its complex conjugate, which we will denote
here by $\bar \chi(r)$, must be considered as formally independent
variables, which can be analytically continued separately, and do not
necessarily satisfy the ``reality condition'' $\bar\chi(r)=\chi(r)^*$.
The same applies to $\phi(r)$ and $\bar\phi(r)$.

\begin{figure}
\centerline{
\epsfxsize=100mm
\epsfbox{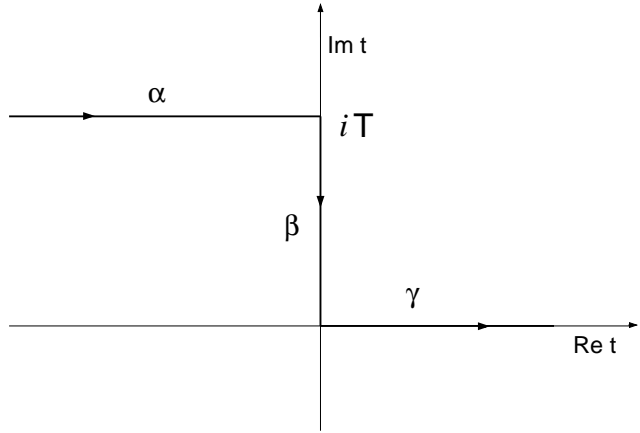}
}
\caption{\tenrm
Complex time contour.
}
\end{figure}

The boundary conditions are of special importance.  The solution
must be real, in the sense specified above,
along $\gamma$, where the asymptotic fields
for $t \to +\infty$ represent the final
state.   The condition on the particle
number in the initial state translates instead into the requirement
that for very early times in the evolution (i.e.~asymptotically
for $t \to -\infty + \imath T$ along $\alpha$)
the fields must reduce to a superposition of normal
modes of oscillation with amplitudes satisfying the equation
\begin{equation}
\bar a(- k) = e^{\theta} a^*( k)
\label{eq43}
\end{equation}
where $k$ is an index characterizing the radial momentum of the waves,
and $\theta$ plays the role of a chemical potential conjugate
to the particle number in the initial state.  It is clear from
Eq.~\ref{eq43} that for $\theta \ne 0$ the fields cannot be real
along $\alpha$.

Computationally, it is convenient to think again of a time reversed evolution
by which, starting from an initial configuration at $t=0$ the fields
undergo a Euclidean evolution along the imaginary time axis
to $t=\imath T$ (i.e. following the oriented segment $-\beta$)
and then a Minkowski evolution (along $-\alpha$) from $\imath T$
to $-\infty + \imath T$.
For $t=0$ for an $x$-axis discretized with $N$ sites
we have as free variables $N$ real values of the field  and $N$ pure
imaginary values of the conjugate momentum per each independent
canonically conjugate field-momentum pair.  The conditions that
the normal mode amplitudes must
satisfy Eq.~\ref{eq43} amount to $N$ complex equations, i.e.~$2N$ real
equations again for each canonically conjugate field-momentum pair.
Thus, in principle,
one could evolve the fields from an initial {\it Ansatz} at $t=0$ and
adjust the initial variables so that Eq.~\ref{eq43} is satisfied.
In practice, since the evolution equations along the imaginary time
axis are elliptic, one cannot perform a forward integration.  Rather,
one must resort to some relaxation procedure or other
global algorithm, by which one solves the evolution
equations as a set of simultaneous non-linear equations
for all points of a space-time grid.  The situation is further
complicated by the fact that, with complexified fields,
one cannot just minimize a Euclidean action integral.
We have developed a ``second order'' formulation, by
which we minimize a constraint functional obtained from
the modulus squared of the functions that must vanish
at all grid points (an earlier part of this study was done in
collaboration with Timothy Vaughan).  In this case also
we have used the formalism of lattice gauge theory to
obtain a gauge invariant discretization. We tested our
procedure in the context of the $2D$ Abelian Higgs model
(one space, one time dimensions), where we found that it
did reproduce the expected Euclidean solutions, including
solutions with multiple bounces of the fields between two
different topological sectors.

Another crucial component of the calculation consists in
solving the evolution equations along the semiaxis $-\alpha$,
from $\imath T$ to $-\infty + \imath T$ and extracting the normal
mode amplitudes $a(k), \bar a(k)$.  Here we feel that we have
formalism and algorithm already in place, although we need to
check that the integration remains stable with complexified fields.

In conclusion, the study of topology changing processes below
the sphaleron barrier presents some additional challenges, the
most notable being the need to integrate elliptic, rather than
hyperbolic, equations over part of complex time path.  The number
of degrees of freedom and the complexity of the calculation,
however, are not substantially different from those which
characterize the numerical investigations of classically allowed
processes, where, as we have shown above, the power of the
computational tools are well adequate to produce interesting
and accurate results.  This warrants the expectation that even
the classically forbidden processes will be amenable to a
successful computational analysis.

{\bf Acknowledgments}

This research was supported
in part under DOE grant DE-FG02-91ER40676 and NSF grant
ASC-940031.  We wish to thank V.~Rubakov for very interesting
conversations which stimulated the investigation described here,
A.~Cohen, K.~Rajagopal and P.~Tinyakov for valuable discussions,
and T.~Vaughan for participating in an early stage of this work.

\end{document}